%% file: post.tex
\documentclass{article}[11pt]
\usepackage{amsmath,amsfonts,amsthm,times}

\input{prestuff}

\def\defeq{\stackrel{\mathrm{def}}{=}}

\newcommand{\ceiling}[1]{\lceil#1\rceil}

\def\Span#1{\textbf{Span}\left(#1  \right)}
\def\nullspace#1{\textbf{Null}\left(#1  \right)}

\def\wdilation#1#2{\textbf{wd}_{#1}\left(#2  \right)}
\def\wcong#1#2{\textbf{wc}_{#1}\left(#2  \right)}

\def\path#1#2{\textbf{path}_{#1}\left(#2  \right)}

\def\bigO#1{O\left(#1  \right)}

\def\norm#1{\left\| #1 \right\|}

\def\setof#1{\left\{#1  \right\}}
\def\sizeof#1{\left|#1  \right|}

\def\setminus{-}

\newcommand\bb{\boldsymbol{\mathit{b}}}
\newcommand\cc{\boldsymbol{\mathit{c}}}

\newcommand\qq{\boldsymbol{\mathit{q}}}
\newcommand\rr{\boldsymbol{\mathit{r}}}
\newcommand\sst{\boldsymbol{\mathit{\tilde{s}}}}

\newcommand\xx{\boldsymbol{\mathit{x}}}
\newcommand\xxt{\boldsymbol{\mathit{\tilde{x}}}}

\newcommand\yy{\boldsymbol{\mathit{y}}}

\newcommand\yyt{\boldsymbol{\mathit{\tilde{y}}}}
\newcommand\zz{\boldsymbol{\mathit{z}}}
\newcommand\zzero{\boldsymbol{\mathit{0}}}

\def\bvec#1{{\mbox{\boldmath $#1$}}}

\def\Span#1{\mbox{{\bf Span}}\left(#1  \right)}
\def\abs#1{\left|#1  \right|}

\newdimen\pIR
\newcommand\StevesR{{\rm I\kern\pIR R}}
\def\Reals#1{\StevesR^{#1}}

\def\eindex#1{\textrm{class}\left(#1\right)}

\begin{document}

\title{
Solving Sparse, Symmetric, Diagonally-Dominant
 Linear Systems in Time $O (m^{1.31})$} 

\author{
Daniel A. Spielman 
\thanks{Partially supported by NSF grant CCR-0112487.  
\texttt{spielman@math.mit.edu}}\\ 
Department of Mathematics\\ 
Massachusetts Institute of Technology
\and 
Shang-Hua Teng 
\thanks{
Partially supported by NSF grant CCR-9972532. \texttt{steng@cs.bu.edu}}\\
Department of Computer Science\\
Boston University and\\
Akamai Technologies Inc.}

\maketitle

\begin{center}
\large
Second post-FOCS revision.
\end{center}

\begin{abstract}
\textit{We present a linear-system solver that, 
  given an $n$-by-$n$ symmetric positive semi-definite, diagonally dominant 
  matrix $A$ with $m$ non-zero entries and an $n$-vector $\bb $,
  produces a vector $\xxt$ within relative distance $\epsilon$
  of the solution to $A \xx = \bb$
  in time $O (m^{1.31} \log (n \kappa_{f} (A)/\epsilon )^{O (1)} )$,
  where $\kappa_{f} (A)$ is the log of the ratio of the largest to smallest non-zero
  eigenvalue of $A$.
In particular, $\log (\kappa_{f} (A)) = O (b \log n)$, where
  $b$ is the logarithm of the ratio of the largest to smallest
  non-zero entry of $A$.
If the graph of $A$ has genus $m^{2\theta }$
  or does not have a $K_{m^{\theta }} $ minor,
  then the exponent of $m$ can be improved to
  the minimum of
  $1 + 5 \theta $ and $(9/8) (1+\theta )$.
The key contribution of our work is an extension
  of Vaidya's techniques for constructing and
  analyzing combinatorial preconditioners.
}\end{abstract}

\input{intro}

\input{support}

\input{subgraphs}

\input{algs}

\bibliographystyle{alpha}
\bibliography{stoccg}

\end{document}

%% file: prestuff.tex
\newtheorem{theorem}{Theorem}[section]

\newtheorem{lemma}[theorem]{Lemma}

\newtheorem{remark}[theorem]{Remark}
\newtheorem{proposition}[theorem]{Proposition}
\newtheorem{definition}[theorem]{Definition}



%% file: intro.tex
\section{Introduction}


Sparse linear systems are ubiquitous in scientific computing 
  and optimization.
In this work, we develop fast algorithms for solving
  some of the best-behaved linear systems: those specified
  by symmetric, diagonally dominant matrices
  with positive diagonals.
We call such matrices PSDDD as they are positive semi-definite
  and diagonally dominant.
Such systems arise in the solution
  of certain elliptic differential equations via
  the finite element method, 
  the modeling of resistive networks,
  and in the
  solution of certain network optimization 
  problems~\cite{StrangFix,Multigrid,Iterative,Iterative2,Iterative3}.

While one is often taught to solve a linear system $A \xx = \bb $
  by computing $A^{-1}$ and then multiplying $A^{-1}$ by $\bb$,
  this approach is quite inefficient for
  sparse linear systems---the best known bound on 
  the time required to compute $A^{-1}$ 
  is $O (n^{2.376})$~\cite{CoppersmithWinograd} and
  the representation of $A^{-1}$ typically requires
  $\Omega (n^{2})$ space.
In contrast, if $A$ is symmetric and has $m$ non-zero entries, then
  one can use the Conjugate Gradient method, as a direct method,
   to solve for
  $A^{-1} \bb$ in $O (nm)$ time and $O (n)$ space!
Until Vaidya's revolutionary introduction of
  combinatorial preconditioners~\cite{Vaidya},
  this was the best complexity bound for the solution
  of general PSDDD systems.

The two most popular families of methods for solving
  linear systems are the direct methods and the
  iterative methods.
Direct methods, such as Gaussian elimination,
  perform arithmetic operations that produce $\xx$
  treating the entries of $A$ and $\bb$ symbolically.
As discussed in Section~\ref{sec:direct}, direct methods
  can be used to quickly compute $\xx$ if the matrix
  $A$ has special topological structure.

Iterative methods, which are discussed
  in Section~\ref{sec:iterative},
  compute successively better approximations
  to $\xx$.
The Chebyshev and Conjugate Gradient methods take
  time proportional to $m \sqrt{\kappa_{f} (A)} \log (\kappa_{f} (A) / \epsilon)$
  to produce approximations to $\xx$ with relative error $\epsilon$,
 where $\kappa_{f} (A)$ is the ratio of the largest to the smallest
  non-zero eigenvalue of $A$.
These algorithms are improved by preconditioning---essentially solving
  $B^{-1} A \xx = B^{-1} \bb $
  for a \textit{preconditioner} $B$ that is carefully chosen
  so that $\kappa_{f} (A, B)$ is small and
  so that it is easy to solve
  linear systems in $B$.
These systems in $B$ may be solved using direct methods,
  or by again applying iterative methods.

Vaidya~\cite{Vaidya} discovered that for PSDDD matrices $A$
  one could use combinatorial techniques to construct matrices
  $B$ that provably satisfy both criteria.
In his seminal work, Vaidya shows that 
  when $B$ corresponds to a subgraph of the graph 
  of $A$,  one can bound
  $\kappa_{f} (A, B)$ by bounding the dilation and congestion
  of the best embedding of the graph of $A$ into the
  graph of $B$.
By using preconditioners derived by
  adding a few edges to maximum spanning trees, Vaidya's algorithm
  finds $\epsilon$-approximate solutions to 
  PSDDD linear systems of maximum valence $d$ in
  time $O ((d n)^{1.75} \log (\kappa_{f} (A) / \epsilon ))$.
\footnote{For the reader unaccustomed to condition numbers,
  we note that for an PSDDD matrix $A$ in which each entry is
  specified using $b$ bits of precision, 
  $\log (\kappa_{f} (A)) = O (b \log n)$.}
When these systems have special structure, such as having a
  sparsity graph of bounded genus or avoiding certain minors,
  he obtains even faster algorithms.
For example, his algorithm solves planar linear systems
  in time $O ((d n)^{1.2} \log (\kappa_{f} (A) / \epsilon ))$.
This paper follows the outline established by Vaidya:
  our contributions are improvements in the techniques
  for bounding $\kappa_{f} (A,B)$, a construction of better
  preconditioners, a construction that depends upon average
  degree rather than maximum degree, and an analysis
  of the recursive application of our algorithm.

As Vaidya's paper was never published%
\footnote{
Vaidya founded the company
  Computational Applications and System Integration
  (http://www.casicorp.com)
 to market his linear system solvers.},
and his manuscript lacked many
  proofs, the task of formally working out his results fell to others.
Much of its
  content appears in the thesis of his student, Anil Joshi~\cite{Joshi}.
Gremban, Miller  and Zagha\cite{Gremban,GrembanMillerZagha} 
  explain parts of Vaidya's paper as well
  as extend Vaidya's techniques. 
Among other results, they found ways of constructing preconditioners by
  \textit{adding} vertices to the graphs and using separator trees.

Much of the theory behind the application of Vaidya's techniques
  to matrices with non-positive off-diagonals
  is developed in~~\cite{SupportGraph}.
The machinery needed to apply Vaidya's techniques directly
  to matrices with positive off-diagonal elements is developed
  in~\cite{MWB}.
The present work builds upon an algebraic extension of the
  tools used to prove bounds on $\kappa_{f} (A, B)$
  by Boman and Hendrickson~\cite{SupportTheory}.
Boman and Hendrickson~\cite{BomanHendricksonAKPW}
  have pointed out that by applying one of their bounds on 
  support to 
  the tree constructed by Alon, Karp, Peleg, and West \cite{AKPW}
  for the $k$-server problem, one obtains
  a spanning tree preconditioner $B$ with 
  $\kappa_{f} (A, B) = m 2^{\bigO{\sqrt{\log n\log\log n}}}$.
They thereby obtain a solver for 
  PSDDD systems that produces $\epsilon $-approximate solutions in
  time $m^{1.5 + o (1)} \log (\kappa_{f} (A) / \epsilon )$.
In their manuscript, they asked whether one could possibly augment
  this tree to obtain a better preconditioner. 
We answer this question in the affirmative.
An algorithm running in time $O (m n^{1/2} \log^{2} (n))$
  has also recently been obtained by Maggs,
  \textit{et. al.}~\cite{MaggsEtAl}.

The present paper is the first to push past the $O (n^{1.5})$ barrier.
It is interesting to observe that this is exactly the point
  at which one obtains sub-cubic time algorithms for solving
  dense PSDDD linear systems.

Reif~\cite{Reif} proved that by applying Vaidya's techniques
  recursively, one can solve bounded-degree planar
  positive definite diagonally dominant linear systems
  to relative accuracy $\epsilon$ in time
  $O (m^{1 + o (1)} \log (\kappa (A) / \epsilon ))$.
We extend this result to general planar PSDDD linear systems.

Due to space limitations in the FOCS proceedings, some proofs have
  been omitted.
These are being gradually included in the on-line version of the paper.

\subsection{Background and Notation}
A symmetric matrix $A$ is semi-positive definite 
  if $x^{T} A x \geq 0$ for all vectors $x$.
This is equivalent to having all eigenvalues of $A$
  non-negative.


In most of the paper, we will focus on Laplacian matrices:
  symmetric 
  matrices with non-negative diagonals and non-positive off-diagonals
  such that for all $i$,  $\sum_{j} A_{i,j} = 0$.
However, our results will apply to the more general family
  of positive semidefinite, diagonally dominant (PSDDD) matrices,
  where a matrix is diagonally dominant if
  $\abs{A_{i,i}} \geq \sum_{j=1}^{n} \abs{A_{i,j}}$ for all $i$.
We remark that a symmetric matrix is PSDDD if and only if 
  it is diagonally dominant and all of its diagonals are
  non-negative.

In this paper, we will restrict our attention to the solution
  of linear systems of the form $A \xx = \bb$
  where $A$ is a PSDDD matrix.
When $A$ is non-singular, that is when $A^{-1}$ exists,
  there exists a unique solution $x = A^{-1}\bb $ to the linear
  system.
When $A$ is singular and symmetric, 
  for every $\bb  \in \Span{A}$ there exists a unique
  $\xx \in \Span{A}$ such that $A \xx = \bb$.
If $A$ is the Laplacian of a connected graph,
  then the null space of $A$ is spanned by $\bvec{1}$.

There are two natural ways to formulate the problem of finding
  an approximate solution to a system $A \xx = \bb$.
A vector $\xxt$ has \textit{relative residual error} $\epsilon$
  if $\norm{A \xxt - \bb} \leq \epsilon \norm{\bb }$.
We say that a solution $\xxt$ is an $\epsilon$-approximate
  solution if it is at relative
  distance at most $\epsilon$ from the actual
  solution---that is, if 
  $\norm{\xx - \xxt  } \leq \epsilon \norm{\xx }$.
One can relate these two notions of approximation by observing that
  relative distance of $\xx$ to the solution and
  the relative residual error differ by a multiplicative
  factor of at most $\kappa_{f} (A)$.
We will focus our attention on the problem
  of finding $\epsilon$-approximate solutions.
 
The ratio $\kappa_{f} (A)$ is the finite condition number of $A$.
The $l_{2}$ norm of a matrix, $\norm{A}$, is the maximum of
  $\norm{ A x} / \norm{x}$, and equals the largest eigenvalue
  of $A$ if $A$ is symmetric.
For non-symmetric matrices,
  $\lambda_{max} (A)$ and $\norm{A}$ are typically different.
We let $|A|$ denote the number of non-zero entries in $A$, and
  $\min (A)$ and $\max (A)$ denote the smallest and largest
  non-zero elements of $A$ in absolute value, respectively.

The condition number plays a prominent role in the analysis
  of iterative linear system solvers.
When $A$ is PSD, it is known that, after
  $\sqrt{\kappa_{f} (A)} \log (1/\epsilon )$ iterations,
  the Chebyshev iterative method and the Conjugate Gradient method
  produce solutions with relative residual error at most $\epsilon$.
To obtain an $\epsilon$-approximate solution, one need merely
  run $\log (\kappa_{f} (A))$ times as many iterations.
If $A$ has $m$ non-zero entries, each of these iterations takes
  time $O (m)$.
When applying the preconditioned versions of these algorithms
  to solve systems of the form $B^{-1} A \xx = B^{-1} \bb $,
  the number of iterations required by these algorithms 
  to produce an $\epsilon$-accurate solution is bounded
  by 
  $\sqrt{\kappa_{f} (A, B)} \log (\kappa_{f} (A) /\epsilon ) $
  where 
\[
  \kappa_{f} (A, B) 
= 
\left(\max_{\xx : A\xx \neq \zzero} \frac{ \xx^{T} A \xx}{\xx^T B \xx}
 \right)
\left(\max_{\xx : A\xx \neq \zzero} \frac{ \xx^{T} B \xx}{\xx^T A \xx}
 \right),
\]
for symmetric $A$ and $B$ with $\Span{A} = \Span{B}$.
However, each iteration of these methods takes time
  $O (m)$ plus the time required to solve linear
  systems in $B$.
In our initial algorithm, we will use direct methods to
  solve these systems, and so will not have to worry about
  approximate solutions.
For the recursive application of our algorithms, we will
  use our algorithm again to solve these systems, and so will
  have to determine how well we need to approximate the solution.
For this reason, we will analyze the Chebyshev iteration instead
  of the Conjugate Gradient, as it is easier to analyze the impact
  of approximation in the Chebyshev iterations. 
However, we expect that similar results could be obtained for
  the preconditioned Conjugate Gradient.
For more information on these methods, we refer the reader
  to \cite{GolubVanLoanBook} or \cite{Bruaset}.

\subsection{Laplacians and Weighted Graphs}
All weighted graphs in this paper have
  positive weights.
There is a natural isomorphism between weighted
  graphs and Laplacian matrices:
  given a weighted graph $G = (V, E, w)$, we can
  form the Laplacian matrix in which
  $A_{i,j} = -w (i,j)$ for $(i,j) \in E$,
  and with diagonals determined by the condition
  $A \bvec{1} = \bvec{0}$.
Conversely, a weighted graph is naturally associated
  to each Laplacian matrix.
Each vertex of the graph corresponds to both a row and
  column of the matrix, and we will often
  abuse notation by identifying this row/column pair
  with the associated vertex.

We note that if $G_{1}$ and $G_{2}$ are weighted 
  graphs on the same vertex set with disjoint sets
  of edges, then the Laplacian of the union of
  $G_{1}$ and $G_{2}$ is the sum of their
  Laplacians.

\subsection{Reductions}\label{sec:reductions}

In most of this paper we just consider
  Laplacian matrices of connected graphs.
This simplification is enabled by two reductions.

First, we note that it suffices to construct preconditioners
  for matrices satisfying
  $A_{i,i} = \sum_{j}\abs{A_{i,j}}$, for all $i$.
This follows from the observation in~\cite{SupportGraph}
  that if $\tilde{A} = A + D$, where
  $A$ satisfies the above condition, then
  $\kappa _{f} (\tilde{A}, B + D) \leq \kappa _{f} (A,B)$.
So, it suffices to find a preconditioner after 
  subtracting off the maximal diagonal matrix that maintains
  positive diagonal dominance.

We then use an idea of Gremban~\cite{Gremban} for handling
  positive off-diagonal entries.
If $A$ is a symmetric matrix such that for all $i$,
  $A_{i,i} \geq  \sum _{j} \abs{A_{i,j}}$,
  then Gremban decomposes 
  $A$ into $D + A_{n} + A_{p}$, where
  $D$ is the diagonal of $A$,
  $A_{n}$ is the matrix containing all 
  negative off-diagonal entires of $A$,
  and $A_{p}$ contains all the positive off-diagonals.
Gremban then considers the linear system
\[
  \left[
\begin{array}{ll}
  D + A_{n} & -A_{p}\\
  -A_{p} & D + A_{n}
\end{array}
 \right]
\left[
\begin{array}{l}
\xx\\
\xx'
\end{array}
 \right]
=
\left[
\begin{array}{l}
\bb\\
-\bb
\end{array}
 \right],
\]
and observes that its solution will have
  $\xx' = -\xx$ and that 
  $\xx$ will be the solution to 
  $A \xx = \bb $.
Thus, by making this transformation,
  we can convert any $PSDDD$ linear
  system into one with 
  non-negative off diagonals.
One can understand this transformation as
  making two copies of every vertex in the graph,
  and two copies of every edge.
The edges corresponding to negative off-diagonals
  connect nodes in the same copy of the graph,
  while the others cross copies.
To capture the resulting family of graphs, we
  define a weighted graph $G$ to be a
  \textit{Gremban cover}
  if it has $2n$ vertices and
\begin{itemize}
\item for $i,j \leq n$, 
  $(i,j) \in  E$ if and only if
  $(i+n, j+n) \in E$, and
  $w (i,j) = w (i+n, j+n)$, 
\item for $i,j \leq n$, 
  $(i,j+n) \in  E$ if and only if
  $(i+n, j) \in E$, and
  $w (i,j+n) = w (i+n, j)$, and
\item the graph contains no edge
  of the form $(i, i+n)$.
\end{itemize}
When necessary,
  we will explain how to modify our arguments
  to handle Laplacians that are Gremban covers.

Finally, if $A$ is the Laplacian of an unconnected
  graph, then the blocks corresponding to the connected
  components may be solved independently.

\subsection{Direct Methods}\label{sec:direct}
The standard direct method for solving symmetric linear systems
  is Cholesky factorization.
Those unfamiliar with Cholesky factorization should think of
  it as Gaussian elimination in which one 
  simultaneously eliminates on rows and columns so as to preserve
  symmetry.
Given a permutation matrix $P$,
  Cholesky factorization produces a lower-triangular matrix
  $L$ such that $L L^{T} = P A P^{T}$.
Because one can use forward and back substitution to
  multiply vectors by $L^{-1}$ and $L^{-T}$
  in time proportional to the
  number of non-zero entries in $L$,
  one can use the Cholesky factorization of $A$
  to solve the system
  $A \xx = \bb $ in time $O (\sizeof{L})$.
 
Each pivot in the factorization comes from the diagonal
  of $A$, and one should understand the 
  permutation $P$ as providing
  the order in which these pivots are chosen.
Many heuristics exist for producing permutations $P$ 
  for which the number of non-zeros in $L$ is small.
If the graph of $A$ is a tree, then a permutation
  $P$ that orders the vertices of $A$ from the leaves up
  will result in an $L$ with at most $2n-1$ non-zero entries.
In this work, we will use results concerning matrices
  whose sparsity graphs resemble trees with a few additional
  edges and whose graphs have small separators, which
  we now review.

If $B$ is the Laplacian matrix of a weighted graph
  $(V,E,w)$, and one eliminates a vertex $a$
  of degree $1$, then the remaining matrix
  has the form
\[
  \left[
  \begin{array}{ll}
  1 & 0\\
  0 & A_{1},
\end{array}
 \right]
\]
where $A_{1}$ is the Laplacian of the graph in which    
  $a$ and its attached edge have been removed.
Similarly, if a vertex $a$ of degree $2$ is eliminated,
  then the remaining matrix is the Laplacian of the   
  graph in which the vertex $a$ and its adjacent edges
  have been removed, and an edge
  with weight $1/ (1/w_{1} + 1/w_{2})$
  is added between the
  two neighbors of $a$,
  where $w_{1}$ and $w_{2}$ are the weights of the edges
  connecting $a$ to its neighbors.

Given a graph $G$ 
  with edge set $E = R \cup S$, where the edges
  in $R$ form a tree,
  we will perform a partial Cholesky factorization
  of $G$ in which we successively eliminate all the degree 1 and 2
  vertices that are not endpoint of edges in $S$.
We introduce the algorithm \texttt{trim} to define the order
  in which the vertices should be eliminated, and we call the
  \emph{trim order} the order in which \texttt{trim} deletes vertices.
\begin{trivlist}
\item []
\noindent {\bf Algorithm:} \texttt{trim$(V,R,S)$}
\begin{enumerate}
\item  While $G$ contains a vertex of degree one
  that is not an endpoint of an edge in $S$,
  remove that vertex and its adjacent edge.
\item  While $G$ contains a vertex of degree two
  that is not an endpoint of an edge in $S$,
  remove that vertex and its adjacent edges, and add
  an edge between its two neighbors.
\end{enumerate}
\end{trivlist}

\begin{proposition}\label{pro:trim}
The output of \texttt{trim} is a graph
  with at most $4 \sizeof{S}$ vertices
  and $5 \sizeof{S}$ edges.
\end{proposition}
 
\begin{remark}
If $(V,R)$ and $(V,S)$ are Gremban covers,
  then we can implement \texttt{trim} so
  that the output graph is also a Gremban cover.
Moreover, the genus and maximum size clique minor
  of the output graph do not increase.
\end{remark}

After performing partial Cholesky factorization
  of the vertices in the trim order, one obtains
  a factorization of the form
\[
B = L C
    L^{T},
\mbox{where $C = $ }
\left[
    \begin{array}{ll}
     I & 0\\
     0 & A_{1}
    \end{array}
 \right],
\]
$L$ is lower triangular, 
  and the left column and right columns in the above
  representations correspond to the eliminated 
  and remaining vertices
  respectively.
Moreover, $\sizeof{L} \leq 2n-1$, and
  this Cholesky factorization may be performed in
  time $O (n + \sizeof{S})$.


The following Lemma may be proved by induction.
\begin{lemma}\label{lem:partialCholesky}
Let $B$ be a Laplacian matrix and let
  $L$ and $A_{1}$ be the matrices arising from
  the partial Cholesky factorization of $B$
  according to the trim order.
Let $U$ be the set of eliminated vertices, 
  and let $W$ be the set of remaining vertices.
For each pair of vertices $(a,b)$ in $W$ joined by a simple
  path containing only vertices of $U$, let
  $B_{(a,b)}$ be the Laplacian of the graph containing
  just one edge between $a$ and $b$ of weight
  $1/(\sum_{i} 1/w_{i})$, where
  the $w_{i}$ are the weights on the path between
  $a$ and $b$.
Then, 

\begin{itemize}
\item [$(a)$]
  the matrix $A_{1}$ is the sum of the Laplacian of the 
  induced graph
  on $W$ and
  the sum all the Laplacians $B_{(a,b)}$, 
\item [$(b)$]
  $\norm{A_{1}} \leq \norm{B}$,
  $\lambda_{2} (A_{1}) \geq \lambda_{2} (B)$, and so
  $\kappa _{f} (A_{1}) \leq \kappa _{f} (B)$.
\end{itemize}
\end{lemma}

Other topological structures may be exploited
  to produce elimination orderings
  that result in sparse $L$.
In particular,  Lipton, Rose and Tarjan~\cite{LiptonRoseTarjan} 
  prove that if the sparsity graph is 
  planar, then one can find such an $L$ with at most
  $O (n \log n)$ non-zero entries in time $O (n^{3/2})$.
In general, Lipton, Rose and Tarjan prove that if
  a graph can be dissected by a family of small separators,  
  then $L$ can be made sparse.
The precise definition and theorem follow.
  
\begin{definition}\label{def:separator}
A subset of vertices $C$ of a graph $G= (V,E)$ with $n$ vertices is an 
  $f (n)$-separator if 
  $\sizeof{C}\leq f (n)$, and the vertices of $V - C$
  can be partitioned into two sets  $U$ and $W$ such that there are
  no edges from $U$ to $W$, and $\sizeof{U},\sizeof{W}\leq 2n/3 $.
\end{definition}

\begin{definition}\label{def:familyseparator}
Let $f ()$ be a positive function.
A graph $G = (V,E)$ with $n$ vertices has a family of $f ()$-separators
  if for every $s \leq  n$, every subgraph $G' \subseteq G$ with $s$ vertices
  has a $f (s)$-separator.
\end{definition}

\begin{theorem}[Nested Dissection: Lipton-Rose-Tarjan]
  \label{thm:nestesdissection}
Let $A$ be an $n$ by $n$ symmetric PSD matrix, $\alpha > 0$ be a
  constant, and $h (n)$ be a positive function of $n$.
Let $f (x) = h (n) x^{\alpha }$.
If $G (A)$ has a family of $f ()$-separator, then 
  the Nested Dissection Algorithm of Lipton, Rose and Tarjan can,
 in $\bigO{n + (h (n) n^{\alpha })^{3}}$ time,
 factor $A$ into $A = LL^{T}$ 
 so that $L$ has at most $\bigO{(h (n)n^{\alpha })^{2}\log n}$
  non-zeros.
\end{theorem}

To apply this theorem, we note that many families of graphs
  are known to have families of small separators.
Gilbert, Hutchinson, and Tarjan
  \cite{GilbertHutchinsonTarjan} show
  that all graphs of $n$ vertices
  with genus bounded by $g$ have a family of $O (\sqrt{gn})$-separators,
  and Plotkin, Rao and Smith \cite{PlotkinRaoSmith} 
  show that any graph that excludes
  $K_{s}$ as minor has a family of $O (s\sqrt{n\log n})$-separators.

\subsection{Iterative Methods}\label{sec:iterative}
Iterative methods such as
  Chebyshev iteration and Conjugate Gradient
  solve systems such as
  $A \xx = \bb $
  by successively multiplying vectors
  by the matrix $A$, and
  then taking linear combinations of
  vectors that have been produced so far.
The preconditioned versions of these
  iterative methods take as input
  another matrix $B$, called the
  \textit{preconditioner},
  and also perform
  the operation of solving linear systems
  in $B$.
In this paper, we will restrict our attention to 
  the preconditioned Chebyshev method as it
  is easier to understand the effect of
  imprecision in the solution of the systems in 
  $B$ on the method's output.
In the non-recursive version of our algorithms,
  we will exploit the standard analysis
  of Chebyshev iteration (see~\cite{Bruaset}), adapted
  to our situation:

\begin{theorem}[Preconditioned Chebyshev]\label{thm:cheby}
Let $A$ and $B$ be Laplacian matrices, let $\bb $
  be a vector, and let $\xx$ satisfy $A \xx  = \bb$.
At each iteration, the preconditioned Chebyshev method
  multiplies one vector by $A$, solves one linear system
  in $B$, and performs a constant number of vector additions.
At the $k$th iteration, the algorithm maintains a solution
  $\xxt$ satisfying
\[
  \norm{(\xxt - \xx)}
 \leq 
    e^{-k / \sqrt{\kappa_{f} (A,B)}} 
   \kappa_{f} (A) \sqrt{\kappa_{f} (B)} \norm{\xx}.
\]
\end{theorem}

In the non-recursive versions of our algorithms,
  we will pre-compute the Cholesky factorization of 
  the preconditioners $B$,
  and use these to solve the linear systems encountered
  by preconditioned Chebyshev method.
In the recursive versions, we will perform a
  partial Cholesky factorization of $B$,
  into a matrix of the form $L [I, 0; 0, A_{1}] L^{T}$ ,
  construct a preconditioner for $A_{1}$, and again use
  the preconditioned Chebyshev method to solve
-  the systems in $A_{1}$.



%% file: support.tex
\section{Support Theory}

The essence of support theory is the realization that one
  can bound $\lambda_{f} (A,B)$ by constructing an embedding
  of $A$ into $B$.
We define a \textit{weighted embedding} of $A$ into $B$
  to be a function $\pi$ that maps each edge $e$ of $A$ into a 
  weighted simple path in $B$ linking the endpoints of $A$.
Formally, $\pi : E_{A} \times E_{B} \rightarrow \Reals{+}$
  is a weighted embedding if for all $e \in A$,
  $\setof{f \in B : \pi (e,f) > 0}$ is a simple path connecting
  from one endpoint of $e$ to the other.
We let $\path{\pi}{e}$ denote this set of edges in this path in $B$.
For $e\in A$, we define 
$ \wdilation{\pi}{e}
 =
  \sum_{f \in \path{\pi}{e}} \frac{a_{e}}{b_{f} \pi (e,f)}.$
and the weighted congestion of an edge $f \in B$ under $\pi$ to be
$  \wcong{\pi}{f}
 =
  \sum_{e : f \in \path{\pi}{e}} \wdilation{\pi}{e} \pi (e,f).$


Our analysis of our preconditioners relies on the following
  extension of the support graph theory.

\begin{theorem}[Support Theorem]\label{thm:support}
Let $A$ be the Laplacian matrix of a weighted graph $G$ and $B$
  be the Laplacian matrix of a subgraph $F$ of $G$.
Let $\pi $ be a weighted embedding of $G$ into $F$.
Then 
\[
\kappa_{f} (A,B) 
 \leq
\max_{f \in F} \wcong{\pi}{f}.
\]
\end{theorem}

To understand this statement, the reader should first consider
  the case in which all the weights $a_{e}$, $b_{f}$ and $\pi (e,f)$
  are 1.
In this case, the Support Theorem says that $\kappa_{f} (A,B)$
  is at most the maximum over edges $f$ of the sum of the lengths
  of the paths through $f$.
This improves upon the upper bound on $\kappa_{f} (A,B)$
  stated by Vaidya and proved in Bern \textit{et. al.} of the maximum
  congestion times the maximum dilation, and it improves
  upon the bound proved by Boman and Hendrickson which was the
  sum of the dilations.
This statement also extends the previous theories by using fractions
  of edges in $B$ to route edges in $A$.
That said, our proof of the Support Theorem owes a lot to the machinery
  developed by Boman and Hendrickson and our $\pi$ is analogous to
  their matrix $M$.

We first recall the definition of the support of $A$ in $B$, denoted
  $\sigma (A,B)$:
\[
  \sigma (A, B)
 =
 \min \setof{\tau : \forall t \geq  \tau , \quad 
    t B  \succeq A}.
\]
Gremban proved that one can use support to characterize $\lambda_{f}$:
\begin{lemma}\label{lem:sigma}
If $\nullspace{A} = \nullspace{B}$, then
\[
\lambda_{f} (A, B) = \sigma (A, B) \sigma (B, A).
\]
\end{lemma}
Vaidya observed
\begin{lemma}\label{lem:subgraphSupport}
If $F$ is a subgraph of the weighted graph
  $G$, $A$ is the Laplacian of $G$ and $B$ is the Laplacian of $F$,
  then $\sigma (B,A) \leq 1$.
\end{lemma}
Our proof of the Support Theorem will use the Splitting Lemma
  of Bern \textit{et. al.} and the Rank-One Support Lemma of Boman-Hendrickson:
\begin{lemma}[Splitting Lemma]\label{lem:splitting}
Let $A = A_{1} + A_{2} + \dotsb + A_{k}$ and let
  $B = A_{1} + B_{2} + \dotsb + B_{k}$.
Then,
\[
  \sigma (A, B) \leq \max_{i} \sigma (A_{i}, B_{i}).
\]
\end{lemma}

For an edge $e \in A$ and a weighted embedding $\pi$ of $A$ into $B$,
  we let $A_{e}$ denote the Laplacian of the graph containing only
  the weighted edge $e$ and $B_{e}$ denote the Laplacian of the graph
  containing the edges $f \in \path{\pi}{\epsilon}$ with weights
  $a_{f} \pi (e,f)$.
We have:

\begin{lemma}[Weighted Dilation]\label{lem:rankOne}
For an edge $e \in A$,
\[
  \sigma (A_{e}, B_{e}) = \wdilation{\pi}{e}.
\]
\end{lemma}
\begin{proof}
Follows from Boman and Hendrickson's Rank-One Support Lemma.
\end{proof}

\begin{proof}[Proof of Theorem~\ref{thm:support}]
Lemma~\ref{lem:rankOne} implies
\[
  \sigma (A_{e}, \wdilation{\pi}{e} B_{e}) = 1.
\]
We then have
\begin{align*}
  \sigma (A, \max_{f \in B} \wcong{\pi}{f} B)
 & \leq 
   \sigma (A, \sum_{f \in B} \wcong{\pi}{f} A_{f})\\
 & =
   \sigma (A, \sum_{e \in A} \wdilation{\pi}{e} B_{e})\\
 & \leq
   \max_{e \in A}
   \sigma (A_{e}, \wdilation{\pi}{e} B_{e})\\
 & \leq 1,
\end{align*}
where the second-to-last inequality follows from the Splitting Lemma.
\end{proof}

%% file: subgraphs.tex
\def\calW{\mathcal{W}}


\section{The Preconditioner}\label{sec:supportinggraph}

In this section, we construct and analyze our preconditioner.

\begin{theorem}\label{thm:main}
Let $A$ be a Laplacian and  $G = (V,E,w) $
  its corresponding weighted graph.
Let $G$ have $n$ vertices and $m$ edges.
For any positive integer $t \leq  n$, 
  the algorithm \texttt{precondition}, 
  described below, runs
  in $O (m \log m)$ time and
  outputs a spanning tree $R \subseteq E$ of $G$
  and a set of edges $S \subseteq E$ such that
\begin{itemize}
\item [(1)] if $B$ is the Laplacian corresponding
  to $R \cup S$, then
   $\sigma_{f} (A,B) \leq \frac{m}{t}2^{\bigO{\sqrt{\log n\log\log
  n}}}$, and

\item [(2)] $\sizeof{S} \leq \bigO{t^{2}\log n/\log \log n}$.
\end{itemize}
Moreover, if $G$ has genus $s^{2}$ or has no $K_{s}$ minor, 
  then 
\begin{itemize}
\item [(2')] $\sizeof{S}\leq \bigO{t s \log s \log n / \log \log n}$,
\end{itemize}
and if $G$ is the Gremban cover of such a graph, then
  the same bound holds and
  we can ensure that $S$ is a Gremban cover as well.
\end{theorem}
\begin{proof}
Everything except the statement concerning Gremban
  covers follows immediately from
  Theorem~\ref{thm:support} and
  Lemmas~\ref{lem:sizeEB}, \ref{lem:preconditionTime},
  and~\ref{lem:conjestion}.

In the case that $G$ is Gremban cover, we apply 
  the algorithm \texttt{precondition} to
  the graph that it covers, but keeping all
  weights positive.
We then set $R$ and $S$ to be both images of each
  edge output by the algorithm.
Thus, the size of the set $S$ is at most
  twice what it would otherwise be.

For our purposes, the critical difference between
  these two graphs is that 
  a cycle in the covered graph corresponds 
  in the Gremban cover to
  either two disjoint cycles or a double-traversal of
  that cycle.
Altering the arguments to compensate for
  this change increases the bound
  of Lemma~\ref{lem:dilationbound} 
  by at most a factor of $3$, and the bound
  of Lemma~\ref{lem:conjestion} by at most $9$.
\end{proof}
  
The spanning tree $R$ is built using an algorithm
  of Alon, Karp, Peleg, and West~\cite{AKPW}.
The edges in the set $S$ are constructed by using other
  information generated by this algorithm.
In particular, the AKPW algorithm builds its spanning
  tree by first building a spanning forest, then building
  a spanning forest over that forest, and so on.
Our algorithm works by decomposing the trees in these
  forests, and then adding a representative edge between
  each set of vertices in the decomposed trees.

Throughout this section, we assume without loss of
  generality that the maximum weight of an edge is 1.

\subsection{The Alon-Karp-Peleg-West Tree} \label{sub:AKPWWeighted}

We build our preconditioners by adding edges to the spanning
  trees constructed by  Alon, Karp, Peleg and West \cite{AKPW}.
In this subsection, we review their algorithm, state the properties
  we require of the trees it produces, and introduce the notation
  we need to define and analyze our preconditioner.

The AKPW algorithm is run with
  the parameters
  $x  =  2^{\sqrt{\log n\log\log n}}$ and
  $\rho  =  \ceiling{\frac{3\log n}{\log x}}$,
  and the parameters
  $\mu  =  9\rho \log n$ and
  $y  =  x\mu$ are used in its analysis.

We assume, without loss of generality, that the maximum weight
  edge in $E$ has weight 1.
The AKPW algorithm begins by partitioning the
  edge set $E$ by weight as follows:
\[
E_{i} = \left\{e\in E:  1/ y^{i} < w (e) \leq 1/ y^{i-1} \right\}.
\]
For each edge $e\in E$, let $\eindex{e}$ be the index such that 
  $e\in E_{\eindex{e}}$.

The AKPW algorithm iteratively applies a modification
  of an algorithm of Awerbuch~\cite{Awerbuch}, which
  we call \texttt{cluster}, whose relevant
  properties are summarized in the following lemma.

\begin{lemma}[Colored Awerbuch]\label{lem:coloredawcover}
There exists an algorithm with template
\[
  F = \mathtt{cluster} (G, x, E_{1}, \dotsc , E_{k}),
\]
where $G = (V,E)$ is a graph, $x$ is a number, 
  $E_{1}, \dotsc , E_{k}$ are disjoint subsets of $E$,
  and $F$ is a spanning forest of $V$, 
  such that 
\begin{itemize}
\item [(1)] each forest of $F$ has depth at most $3 x k \log n$,

\item [(2)] for each $1 \leq i \leq k$, the number of edges
   in class $E_{i}$ 
   between vertices in the same tree of $F$
   is at least $x$ times the number of edges in class $E_{i}$ 
   between vertices in distinct trees of $F$, and
\item [(3)] $\mathtt{cluster}$ runs in time
  $O (\sum_{i} \sizeof{E_{i}})$.
\end{itemize}
\end{lemma}

\begin{proof}
Properties $(1)$ and $(2)$ are established in the proof of
  Lemma 5.5 in~\cite{AKPW}.
To justify the running time bound, we review the algorithm.
We first recall that it only pays attention to edges in 
  $\cup_{i} E_{i}$.
The algorithm proceeds by growing a BFS
  tree level-by-level 
  from a vertex that is not included in the current forest.
It grows this tree until a level is reached at which condition
  $(2)$ is satisfied.
Once condition $(2)$ is satisfied, it adds this tree to the forest,
  and begins to grow again from a vertex not currently in the forest.
\end{proof}

The other part of the AKPW algorithm is a subroutine
  with template
\[
   G' = \mathtt{contract} (G, F),
\]
that takes as input a graph $G$ and a spanning forest $F$ 
  of $G$,
  and outputs the multigraph $G'$ obtained by contracting the
  vertices of each tree in $F$ to a single vertex.
This contraction removes all resulting self-loops
  (which result from edges between vertices in the same tree),
  but keeps an image of each edge between distinct trees
  of $F$.
The classes, weights, and names of the edges are preserved,
  so that each edge in $G'$ can be mapped back to a unique
  pre-image in $G$.

We can now state the AKPW algorithm:

\begin{trivlist}
\item []
\noindent {\bf Algorithm:} \texttt{ $R = $ AKPW$(G)$}
\begin{enumerate}
\item  Set $j = 1$ and $G^{(j)} = G$.

\item While $G^{(j)}$ has more than one vertex
\begin{enumerate}
\item 
  Set $R^{j} = \texttt{cluster} (G^{(j)},x,E_{j-\rho +1}, \dots , E_{j})$.

\item  Set $G^{(j+1)} = \mathtt{contract} (G^{(j)}, R^{j})$
\item Set $j = j + 1$.
\end{enumerate}

\item Set $R = \cup_{j} R^{j}$
\end{enumerate}
\end{trivlist}

The tree output by the AKPW algorithm is the union
  of the pre-images of the edges in forests
  $R^{j}$.
Our preconditioner will include these edges, and
  another set of edges $S$ constructed using the 
  forests $F^{j}$.
  
To facilitate the description and analysis of our algorithm, 
  we define
\begin{itemize}
\item [] $F^{j}$ to be the forest on $V$
  formed from the union of the pre-images of
  edges in $R^{1} \cup \dotsb \cup R^{j-1}$,

\item [] $T^{j}_{v}$ to be the tree of $F^{j}$
  containing vertex $v$.

\item [] $E^{j}_{i} 
  = \setof{(u,v) \in E_{i} : T^{j}_{u} \not = T^{j}_{v}}$,

\item [] $H^{j}_{i} = E_{i}^{j} - E_{i}^{j+1}$, and
   $H^{j} = \cup_{i} H^{j}_{i}$.

\end{itemize}
We observe that $F^{j+1}$ is comprised of edges from
  $E_{1},\dotsc ,E_{j}$, and that each edge in
  $H^{j}$ has both endpoints in the same tree of
  $F^{j+1}$.

Alon, \textit{et. al.} prove:

\begin{lemma}[AKPW Lemma 5.4]\label{lem:edgereduction}
The algorithm \texttt{AKPW} terminates. 
Moreover,
  for every $i\leq j$,
$\sizeof{E_{i}^{j}} \leq 
  \sizeof{E_{i}^{(j-1)}}/x \leq 
  \sizeof{E_{i}}/x^{j-i}.$
\end{lemma}

We remark that $x^{\rho } > \sizeof{E}$, so 
  for $i \leq j - \rho $,
  $E_{i}^{j} = \emptyset$.
The following lemma follows from the proof of
  Lemma~5.5 of \cite{AKPW} and the observation that
  $y^{\rho } \geq \sizeof{E}$.
\begin{lemma}\label{lem:pathbound}
For each simple path $P$ in $F^{j+1}$ and for each $l$, 
  $\sizeof{P\cap E_{l}} \leq \min (y^{j-l+1}, y^{\rho})$.
\end{lemma}




\subsection{Tree Decomposition}\label{sec:treedecomp}

Our preconditioner will construct the edge set $S$ by
  decomposing the trees in the forests produced by the
  AKPW algorithm, and adding edges between the resulting sub-trees.
In this section, we define the properties the decomposition
  must satisfy and describe the decomposition algorithm.

\begin{definition}
For a tree $T$ and a set of edges $H$ between the vertices
  of $T$, we define an \textit{$H$-decomposition} of $T$
  to be a pair $(\calW , \sigma )$ where
  $\calW $ is a collection of subsets of the vertices of $T$
  and $\sigma $ is a map from $H$ into 
  sets or pairs of sets in $\calW $
  satisfying
\begin{itemize}
\item [1.] for each set $W \in \calW $, the graph induced
  by $T$ on $W$ is connected,
\item [2.] for each edge in $T$ there is exactly one set
  $W \in \calW $ containing that edge, and
\item [3.] for each edge in $e \in H$, 
  if $\sizeof{\sigma (e)} = 1$, then both endpoints
  of $e$ lie in $\sigma (e)$; otherwise, one endpoint
  of $e$ lies in one set in $\sigma (e)$, and the other
  endpoint lies in the other.
\end{itemize}
\end{definition}

We note that there can be sets $W \in \calW $ containing
  just one vertex of $T$.

For a weighted set of edges $H$ and an
  $H$-decomposition $(\calW , \sigma )$, we define
  the $H$-weight of a set $W \in \calW$ by
$ w_{H} (W) \defeq  \sum _{e \in H : W \in \sigma (e)} w (e)$.

We also define $w_{tot} (H) \defeq \sum _{e \in H} w (e)$.

Our preconditioner will use an algorithm for computing
  small $H$-decompositions in which each set $W \in \calW $
  with $\sizeof{W} > 1$ has bounded $H$-weight.

\begin{lemma}[Tree Decomposition]\label{lem:tree}
There exists an algorithm with template
\[
  (\calW , \sigma ) = \mathtt{decompose} (T, H, \phi  )
\]
that runs in time $O (\sizeof{H} + \sizeof{T})$ and outputs
  an $H$-decomposition $(\calW, \sigma )$ satisfying
\begin{itemize}
\item [1.] for all $W \in \calW $ such that 
  $\sizeof{W} > 1$, $w_{H} (W) \leq \phi $, and
\item [2.] $\sizeof{\calW } \leq 4 w_{tot} (H) / \phi $.
\end{itemize}
\end{lemma}
\begin{proof}
We let $T (v)$ denote the set of vertices in the subtree rooted at $v$,
  and for a set of vertices $W$, let 
  $H (W) = \setof{e \in H : e \cap H \neq  \emptyset}$.
We then define $\bar{w} (v) \defeq H (T (v))$.
Let $v_{0}$ denote the root of the tree.
Our algorithm will proceed as if it were computing
  $\bar{w} (v_{0})$
  via a depth-first traversal of the tree,
  except that whenever it encounters a subtree of
  weight more than $\phi /2$,
  it will place nodes from that subtree into a set in $\calW $
  and remove them from the tree.
There are three different cases which determine how the
  nodes are placed into the set and how $\sigma $ is constructed.

If, when processing a node $v$, the algorithm
  has traversed a subset of the children of $v$, $\setof{v_{1}, \dotsc ,v_{k}}$
  such that $\bar{w} (v_{1}) + \dotsb + \bar{w} (v_{k}) \geq \phi /2 $,
  then a set $W$ is created, all the nodes in 
  $\setof{v} \cup _{i=1}^{k} T (v_{i})$ are placed in $W$, and those
  nodes in $\cup _{i=1}^{k} T (v_{i})$
  are deleted from the tree.
If a node $v$ is encountered
  such that 
  $\phi /2 \leq H (T (v)) \leq \phi  $,
  then  a set $W$ is created, the nodes in 
  $T (v)$ are placed in $W$, and those nodes in $W$
  are deleted from the tree.
In either case, 
  for each node $e \in H (W)$ we set $\sigma (e) = \sigma (e) \cup \setof{W}$.

If a node $v$ is encountered 
  which is not handled by either of the preceeding cases
  and for which $\bar{w} (v) > \phi$,
  then two sets $W_{1} = T (v)$ and $W_{2} = \setof{v}$ are created,
 and those nodes in $T (v)$ are deleted from the tree.
For each edge $e \in H (v)$, 
  $W_{2}$ is added to $\sigma (e)$ and for
  each edge $e \in H (T (v) \setminus \setof{v})$, 
  $W_{1}$ is added to $\sigma (e)$.

When the algorithm finally returns from examining the root,
  all the remaining nodes are placed in a final set,
  and this set is added to $\sigma (e)$ for each edge
  $e \in H$ with endpoints in this set.
The algorithm maintains the invariant 
  that whenever it returns from examining a node $v$,
  it has either deleted $v$, or removed enough
  vertices below $v$ so that
  $\bar{w} (v) < \phi /2 $.
To see that the algorithm produces at most $4 w_{tot}/ \phi $ sets,
  we note that each edge in $H$ can contribute
  its weight to at most two sets, and
  that every time the algorithm forms sets, it either forms one
  set with weight at least $\phi /2 $ or two sets
  with total weight at least $\phi  $.
\end{proof}


\subsection{Constructing the Preconditioner}\label{sec:weightedBridge}
We can now describe our algorithm for constructing the
  preconditioner.
We will defer a discussion of how to efficiently implement
  the algorithm to Lemma~\ref{lem:preconditionTime}.

The algorithm will make use of the parameter
\[
    \theta^{(j)} \defeq 
  \left\{\begin{array}{ll}
     x^{j-1} & \mbox{ if $j \leq \rho $}\\
    x^{\rho }y^{j-\rho -1} & \mbox{otherwise}
    \end{array} \right.
\]

\begin{trivlist}
\item []
\noindent {\bf Algorithm:} \texttt{ $(R,S) = $Precondition$(G)$}
\begin{enumerate}
\item  Run $R = \mathtt{AKPW} (G)$.
  Set $h$ to the number of iterations taken by \texttt{AKPW},
  and record $R^{1}, \dotsc , R^{h}$
  and $H^{1} , \dotsc , H^{h}$.

\item For $j = 1 \mbox{ to } h$
\begin{enumerate}
\item  let $\setof{T_{1},\dotsc ,T_{k}}$  be the set of trees in $F^{j+1}$.
\item for  $i = 1 \mbox{ to } k$
\begin{enumerate}
\item  let $H$ be the subset of edges in $H_{j}$ with endpoints in $T_{i}$

\item  Set $(\setof{W_{1},\dots ,W_{l}} , \sigma )$ to \newline
         $\texttt{decompose} (T_{i}, H, \sizeof{E} / t \theta ^{(j)})$
\item  for each $\mu  \leq \nu  \leq l$, let $a_{\mu ,\nu }$
     be the maximum weight edge in $H$ between $W_{\mu }$ and $W_{\nu}$,
     and add $a_{\mu ,\nu }$ to $S$.
\end{enumerate}

\end{enumerate}
\end{enumerate}
\end{trivlist}


\begin{lemma}\label{lem:sizeEB}
Let $S$ be the set of edges produced by \texttt{Precondition}.
Then, 
\[
\sizeof{S} \leq  8 \rho^{2} t^{2} = 
  \bigO{t^{2} \log n / \log\log n}.
\]
Moreover, if $G$ has no $K_{s}$ minor, then
 $\sizeof{S} = 
\bigO{ t s\log s  \log n / \log\log n}.
$
\end{lemma}
\begin{proof}
Let $c_{j}$ be the total number of
  sets produced by applying \texttt{decompose}
  to the trees in $F^{j+1}$.
We first bound $\sum_{j} c_{j}$. 
We have
\[
  \sum _{j} c_{j} 
  \leq 
  \sum _{j} 4 t \theta^{(j)} w_{tot} (H^{j}) / \sizeof{E}
  =
  \left(  4 t / \sizeof{E} \right)
    \sum _{j} \theta^{(j)} \sum _{i} w_{tot} (H^{j}_{i}).
\]
To bound this sum, we set
\[
  h^{j}_{i} 
 =
  \left\{
  \begin{array}{ll}
   0 & \mbox{if $j < i$,}\\
 \sum _{l \leq i}\sizeof{H^{l}_{i}}
  & \mbox{if $i = j$, and}\\
  \sizeof{H^{j}_{i}} &
  \mbox{if $j > i$}.
\end{array}
 \right.
\]
We observe that Lemma~\ref{lem:edgereduction} implies
  $h^{j}_{i} \leq \sizeof{E_{i}} / x^{j-i}$,
  and $h^{j}_{i} = 0$ for $j \geq i + \rho $.
As $\theta ^{(j)}$ is increasing, we have
\begin{align*}
 \sum _{j} \theta^{(j)} \sum _{i} w_{tot} (H^{j}_{i})
& \leq 
 \sum _{j} \theta^{(j)} \sum _{i = j-\rho +1}^{j} h^{j}_{i} / y^{i-1}\\
& =
 \sum _{i} \sum _{j = i}^{i+\rho -1}  \theta^{(j)} h^{j}_{i} / y^{i-1}\\
& \leq 
 \sum _{i} \sizeof{E_{i}} 
  \sum _{j = i}^{i+\rho -1}  \theta^{(j)} / \left( x^{j-i} y^{i-1} \right)\\
& \leq 
 \sum _{i} \sizeof{E_{i}} 
  \rho \\
& \leq 
  \sizeof{E}
  \rho,
\end{align*}
as $\theta^{(j)} \leq x^{j-i}y^{i-1}$ for $j\leq i+\rho - 1$.
Thus, $\sum_{j} c_{j} \leq  4 \rho t$, and, 
 because we add at most one edge between 
  each pair of these sets, we have
  $\sizeof{S} \leq 8 \rho ^{2} t^{2}$.

As observed by Vaidya, a result of Mader \cite{BollobasMader}
  implies that
  if a graph does not have a complete graph on $s$ vertices
  as a minor, then the average degree of every minor of $G$
  is $O (s\log s)$.
Hence, the number of edges added to $S$ at iteration $j$
  is at most $c_{j} s \log s$, and
  so 
\[
\sizeof{S} 
\leq \sum _{j} c_{j} s \log s
\leq 8 \rho t s \log s. 
\]
Finally, a graph of genus $s^{2}$ does not have
  a $K_{\Theta (s)}$ minor.
\end{proof}

Using the dynamic trees data structure of
  Sleator and Tarjan~\cite{SleatorTarjan},
  we prove:

\begin{lemma}\label{lem:preconditionTime}
If $G$ is a graph with $n$ vertices and $m$ edges,
  then the output of \texttt{precondition} can be 
  produced in $O (m \log m)$ time.
\end{lemma}

\begin{proof}
We first observe that \texttt{AKPW} can be implemented
  to run in time $O (m \log m )$, as each edge
  appears in at most $\rho = O (\log m)$ calls to \texttt{coloredAwerbuch},
  and the contractions can be implemented 
  using standard techniques
  to have
  amortized complexity $O (\log m)$ per node.

As $j$ could be large, it could be impractical for
  the preconditioning algorithm to actually examine the
  entire forest $F^{j+1}$ for each $j$.
To overcome this obstacle, we observe that the
  determination of which edges $a_{\mu ,\nu }$ to include
  in $S$ only depends upon the projection of the 
  sets in the decompositions onto vertices at endpoints
  of edges in $H^{j}$.
That is, rather than passing $(T^{j}, H)$ to 
  \texttt{decompose}, it suffices to pass
  the topological tree induced by restricting
  $T^{j}$ to vertices with endpoints in $H$
  (\textit{i.e.,} with non-essential degree 2 nodes removed).
As this tree has size at most $O (\sizeof{H})$,
  we can implement the algorithm in linear time plus
  the time required to produce these trees.
There are many data structures that allow one to
  dynamically add edges to a tree and,
  for any set of vertices in the tree, to produce
  the induced tree on all least common ancestors
  of those vertices.
For example, one can do this if one can determine
  $(i)$ the nearest common ancestor of any pair of vertices,
  and $(ii)$ which of a pair of vertices comes first
  in an in-order.
The dymanic trees   
  of Sleator and Tarjan \cite{SleatorTarjan}
  enable edge additions and nearest common ancestor queries 
  at an amortized
  cost of $O (\log n)$ each,
  and any algorithm that balances search trees using 
  tree rotations, such as red-black trees, 
  enables one to determine relative order of nodes
  in an in-order at a cost of $O (\log n)$
  per addition and querry.
\end{proof}

\subsection{Analyzing the Preconditioner}\label{sec:analyzePrecond}

We will use weighted embeddings of edges into paths
  in $R \cup S$ to bound the quality of our preconditioners.
The weights will be determined by a function $\tau (j,l)$,
  which we now define to be
\[
\tau (j,l) =\left\{\begin{array}{ll}
1 & \mbox{$j-l< \rho$ } \\
\frac{(j-l-\rho +1)^{2}}{y^{j-l-\rho+1}} & \mbox{Otherwise.}
\end{array}\right.
\]
For each edge
  $e \in H^{j}$ and each edge
  $f \in \path{\pi }{e}$, 
  we will set $\pi (e,f) = \tau (j,\eindex{f})$.
We will construct $\pi $ so as to guarantee
  $\eindex{e} < \eindex{f} + \rho $.

It remains to define the paths over which edges are embedded.
For an edge $e = (u,v)$ in $H^{j}$, 
  if $e \in R \cup S$ then we set $\path{\pi}{e}=e$ and $\pi (e,e)=1$.
Otherwise, we let $T$ be the tree in $F^{j+1}$ containing the
  endpoints of $e$ and let $\sigma $ be the function output by
  \texttt{decompose} on input $T$.
If $\sizeof{\sigma (e)} = 1$, then we let $\path{\pi }{e}$ be the
  simple path in $T$ connecting the endpoints of $e$.
Otherwise, we let $\setof{W_{\nu },W_{\mu }} = \sigma (e)$ and
  let $a_{\nu ,\mu }$ be the edge added between $W_{\nu }$
  and $W_{\mu }$.
We then let $\path{\pi }{e}$ be the concatenation
  of the simple path in $T$ from $u$ to $a_{\nu ,\mu }$,
  the edge $a_{\nu ,\mu }$ and the simple path in $T$
  from $a_{\nu ,\mu }$ to $v$.

The two properties that we require of $\tau $ are encapsulated
  in the following lemma.
\begin{lemma}\label{lem:tau}
\begin{itemize}
\item [$(a)$] For all $j \geq 1$,
$ \sum _{l=1}^{j} 
  \frac{y^{l} \min \left(y^{j-l+1}, y^{\rho } \right)}
       {\tau (j,l)}
 \leq 
   y^{j+1} (\rho +2),
 \mbox{ and}
$
\item [$(b)$] For all $l \geq 1$,
$  \sum _{j \geq l} \tau (j,l)
 \leq 
   (\rho + 1)
$.
\end{itemize}
\end{lemma}

\begin{proof}
The first property follows from 
\begin{align*}
 \sum _{l=1}^{j} 
  \frac{y^{l} \min \left(y^{j-l+1}, y^{\rho } \right)}
       {\tau (j,l)}
 & =
 \sum _{l=1}^{j - \rho} 
  \frac{y^{l} y^{\rho } y^{j-l-\rho +1}}
       {(j - l + \rho  + 1)^{2}}
+
 \sum _{l=j - \rho +1}^{j} 
   y^{l} y^{j-l+1}\\
 & =
 \sum _{l=1}^{j - \rho} 
  \frac{y^{j+1}}
       {(j - l + \rho  + 1)^{2}}
+
 \sum _{l=j - \rho +1}^{j} 
   y^{j+1}\\
 & \leq 
   y^{j+1} (\rho +2),
\end{align*}
as $\sum _{l = 1}^{j-l} (j - l + \rho +1)^{-2} \leq 2$.

The second property follows from $\sum _{i\geq 1} i^{2} y^{-i} \leq 1$,
  which holds because $y$ is greater than 
  the real root of $y^{3} - 4y^{2} + 2y -1$, which is about
  $3.51155$.
\end{proof}

We now derive the upper bound we need on the maximum
  weighted congestion of the embedding $\pi $.

\begin{lemma}\label{lem:dilationbound}
For each $j$ and each simple path
  $P$ in $F^{j+1}$,
\[
\sum_{f\in P} 
  \frac{1}{w (f) \tau (j,\eindex{f})}\leq (\rho +2) y^{j+1}.
\]
\end{lemma}
\begin{proof}
\begin{align*}
\sum_{f\in P} \frac{1}{w (f) \tau (j,\eindex{f})}
  & \leq 
 \sum_{l=1}^{j}\sum_{f\in P\cap E_{l}}
  \frac{1}{w (f) \tau (j,l)}\\
& \leq 
 \sum_{l=1}^{j}
  \frac{\min \left(y^{j-l+1}, y^{\rho } \right)}
       {w (f) \tau (j,l)}\\
& \leq 
 \sum_{l=1}^{j}
  \frac{y^{l} \min \left(y^{j-l+1}, y^{\rho } \right)}
       {\tau (j,l)}\\
& \leq 
 y^{j+1} ( \rho +2)
\end{align*}
where the  third-to-last  inequality follows from Lemma
  \ref{lem:pathbound},
 the second-to-last inequality  follows from $f \in E_{l}$,
  and the last inequality follows from
  Lemma~\ref{lem:tau} $(a)$.
\end{proof}

\begin{lemma}\label{lem:wde}
For each edge $e \in E$,
\[
\wdilation{\pi }{e}
 \leq 
  (2\rho + 5) y^{j+1} w (e).
\]
\end{lemma}
\begin{proof}
Let $e \in H_{i}^{j}$, 
  let $T$ be the forest in $F^{j+1}$
  containing the endpoints of $e$,
  and let $(\calW , \sigma )$ be the output
  of \texttt{decompose} on input $T$.
If $\sizeof{\sigma (e)} = 1$, 
  the $e$ is routed over the simple path in $T$ connecting
  its endpoints, so we can apply Lemma~\ref{lem:dilationbound} 
  to show
\[
\wdilation{\pi }{e}
 \leq 
  (\rho + 2) y^{j+1} w (e).
\]
Otherwise, let $\sigma (e) = \setof{W_{\nu }, W_{\mu }}$,
  and observe that $\path{\pi }{e}$ contains
  two simple paths in $T$
  and the edge $a_{\nu ,\mu }$.
Applying Lemma~\ref{lem:dilationbound} to each of these paths
  and recalling 
  $\eindex{a_{\nu ,\mu }} \leq j$, which implies
  $w (a_{\nu , \mu} ) \geq 1/y^{j}$,
  we obtain
\[
\wdilation{\pi }{e}
 \leq 
  2 (\rho + 2) y^{j+1} w (e) + y^{j} w (e)
 \leq 
   (2 \rho + 5) y^{j+1} w (e).
\]
\end{proof}

\begin{lemma}\label{lem:subconjestion}
 For each $f \in R \cup S$ and for each $j$
\[
\sum_{e \in H^{j}: f \in \path{\pi}{e} }
    \wdilation{\pi}{e} \leq 
   \left(2 \rho + 5 \right) \mu ^{\rho } y^{2} \sizeof{E} / t.
\]
\end{lemma}
\begin{proof}
Let $T$ be the tree in $F^{j+1}$ containing the endpoints of $f$,
  and let $(\calW , \sigma )$ be the output of \texttt{decompose}
  on input $T$.
There are two cases two consider: 
  $f$ can either be an edge of $T$,
  or $f$ can be one of the edges $a_{\nu ,\mu }$.
If $f$ is an edge of $T$, let $W$ be the set
  in $\calW $ containing its endpoints.
Otherwise, if $f$ is one of the edges $a_{\nu ,\mu }$,
  let $W$ be the larger of the sets $W_{\nu }$ or $W_{\mu }$.
If $\sizeof{W_{\nu }} = \sizeof{W_{\mu }} = 1$, then
  the only edge having $f$ in its path is $f$ itself,
  in which case the lemma is trivial.
So, we may assume $\sizeof{W} > 1$.
In either case, each edge $e$ for which
  $f \in \path{\pi }{e}$ must have $W \in \sigma (e)$.
Thus,
\begin{align*}
\sum_{e \in H^{j}: f \in \path{\pi}{e} }
    \wdilation{\pi}{e} 
& \leq 
\sum _{e \in H^{j}: W \in \sigma (e)} 
      \wdilation{\pi}{e} \\
& \leq 
 \sum _{e \in H^{j}: W \in \sigma (e)} w (e) \left(2 \rho + 5 \right) y^{j+1}\\
& = 
 w_{H^{j}} (W) \left(2 \rho + 5 \right) y^{j+1}\\
& \leq 
  \left(2 \rho + 5 \right) y^{j+1} \sizeof{E} / t \theta ^{(j)}\\
& \leq 
  \left(2 \rho + 5 \right) y^{2} \mu ^{\rho } \sizeof{E} / t. \qedhere
\end{align*}
\end{proof}

\begin{lemma}\label{lem:conjestion}
Let $R$, $S$ and $\pi $ be constructed as above.
Then,
\[
\max_{f\in R \cup S}\wcong{\pi}{f} =  
  \frac{m}{t} 2^{\bigO{\sqrt{\log n\log\log n}}}.
\]
\end{lemma}

\begin{proof}
For any edge $f \in R \cup S$, we let $l = \eindex{f}$
  and compute
\begin{eqnarray*}
 \wcong{\pi}{f} & = & \sum_{e \in E: f \in \path{\pi}{e} }
    \wdilation{\pi}{e} \pi (e,f) \\
& = & 
\sum_{j}\sum_{e \in H^{j}: f \in \path{\pi}{e} }
    \wdilation{\pi}{e} \tau (j,l) \\
& \leq  & 
\sum_{j} 
\tau (j,l) \left(2 \rho + 5 \right) \mu ^{\rho } y^{2} \sizeof{E} / t\\
& \leq &
(\rho +1) \left(2 \rho + 5 \right) \mu ^{\rho } y^{2} \sizeof{E} / t,\\
& = &
2^{\bigO{\sqrt{\log n\log\log n}}} \sizeof{E} / t.
\end{eqnarray*}
where the second-to-last inequality follows from Lemma
  \ref{lem:subconjestion},
  the last inequality follows from Lemma~\ref{lem:tau} $(b)$,
  and the last equality follows from
  $\mu ^{\rho } = 2^{\bigO{\sqrt{\log n\log\log n}}}$.
\end{proof}


%% file: algs.tex

\newcommand\sss{\boldsymbol{\mathit{s}}}
\def\approximation#1{\tilde{#1}}

\section{One-Shot Algorithms}\label{sec:}

Our first algorithm constructs
  a preconditioner $B$ for the matrix $A$,
  performs a partial Cholesky factorization
  of $B$ by eliminating the vertices in trim order
  to obtain $B = L [I, 0; 0, A_{1}] L^{T}$,
  performs a further Cholesky factorization of
  $A_{1}$ into $L_{1} L_{1}^{T}$, 
  and applies the preconditioned Chebyshev algorithm.
In each iteration of the preconditioned Chebyshev
  algorithm, we solve the linear systems in $B$
  by back-substitution through the Cholesky factorizations.


\begin{theorem}[One-Shot]\label{thm:oneShot}
Let $A$ be an $n$-by-$n$ PSDDD matrix with
  $m$ non-zero entries.
Using a single application of our preconditioner,
  one can solve the system $A \xx = \bb$
  to relative accuracy $\epsilon $
  in time
$\bigO{m^{18/13 + o (1)} \log (\kappa (A)/ \epsilon )}$.
Moreover, if
if the sparsity graph of $A$ does not contain
  a minor isomorphic to the complete graph on $m^{\theta }$
  vertices,
  or if it has genus at most $m^{2\theta }$, for 
  $\theta < 1/3$,
 then the exponent of $m$ can be reduced to
  $1.125 (1 + \theta ) + o (1)$.
\end{theorem}
\begin{proof}
The time taken by the algorithm is the sum of the time required
  to compute the preconditioner, perform the partial
  Cholesky factorization of $B$,
  pre-process $A_{1}$ (either performing Cholesky factorization
  or inverting it),
  and the product of the number of iterations and the time
  required per iteration.
In each case, we will set $t = m^{\gamma  }$ for some constant
  $\gamma $, and note that the number of iterations will be
  $m^{(1 - \gamma )/2 + o (1)}$, 
  and that the matrix $A_{1}$ will depend on $m^{\gamma }$.

If we do not assume that $A$ has special topological structure,
  then $A_{1}$ is a matrix on $m^{2 \gamma +o (1)}$ vertices.
If we solve systems in $A_{1}$ by Cholesky factorization, then
  it will take time $\bigO{m + m^{6 \gamma +o (1)}}$ 
  to perform the factorization
  and time $\bigO{m +  m^{4 \gamma +o (1)}}$ to solve each system.
So, the total time will be
  $m^{(1 - \gamma )/2 + 4 \gamma + o (1)} + m^{6 \gamma +o (1)}$.
Setting $\gamma = 3/13$, we obtain the first result.


If the graph has genus $\theta^{2}$ or does not have
  a $K_{m^{\theta}}$ minor, or is the Gremban cover
  of such a graph, then
  can apply part $(2')$ of Theorem~\ref{thm:main}.
Thus,
  $A_{1}$ is a matrix on $m^{\gamma + \theta +o (1)}$
  vertices.
In the Gremban cover case, the preconditioner is a Gremban
  cover, and so the partial Cholesky factorization
  can ensure that $A_{1}$ is a Gremban cover as well.
As the Gremban cover of a graph has a similar family
  of separators to the graph it covers, in either
  case we can apply the algorithm of Lipton, Rose
  and Tarjan to obtain the Choleksy factorization of
  $A_{1}$.
By Theorem~\ref{thm:nestesdissection}, with 
  $\alpha = 1/2 + \theta $, the 
  time required to perform the factorization
  will be $\bigO{m + m^{\gamma (3 \theta + 3/2) + o (1)}}$,
  and the time required to solve the system will
  be 
\[
\bigO{m^{(1-\gamma )/2} (m + m^{\gamma (2 \theta + 1) + o (1)} }
= 
\bigO{m^{(1-\gamma )/2 + 1 + o (1) }},
\]
provided $\gamma (2 \theta + 1) \leq 1$.
We will obtain the desired result by setting
  $\gamma = (3 - 9 \theta ) / 4$.
\end{proof}

\section{Recursive Algorithms}\label{sec:recursive}

We now show how to apply our algorithm recursively to improve
  upon the running time of the algorithm presented in
  Theorem~\ref{thm:oneShot}.

For numerical reasons, we will use partial $LDL^{T}$-factorization in this section
  instead of partial Cholesky factorizations.
We remind the reader that the $LDL^{T}$-factorization of a matrix
  $B$ is comprised of a lower-triangular matrix $L$ with ones
  on the diagonal, and a diagonal matrix $D$.
The partial $LDL^{T}$ factorization of a matrix $B_{1}$
  has the form
\[
  B_{1} = 
  L 
  \left(\begin{array}{ll} 
     D & 0 \\
     0 & A_{1}
   \end{array}\right) 
  L^{T},
\]
where $D$ is diagonal $L$ has the form
\[
  L = \left(\begin{array}{ll} L_{1,1} & 0 \\
    L_{2,1} & I\end{array}\right)
\]
and $L_{1,1}$ has 1s on the diagonal.

The recursive algorithm is quite straightforward:
  it first constructs the top-level
  preconditioner $B_{1}$ for matrix $A_{0} = A$.
It then eliminates to vertices of $B_{1}$ in
  the trim order to obtain the partial $LDL^{T}$-factorization
  $B_{1} = L_{1} C_{1}L_{1}^{T}$, where
  $C_{1} =  [D_{1},0;0,A_{1}]$.
When an iteration of the preconditioned Chebyshev algorithm
  needs to solve a linear system in $B_{1}$, 
  we use forward- and backward-substitution to solve
  the systems in 
  $L_{1}$ and $L_{1}^{T}$, but recursively apply 
  our algorithm to solve the linear system in $A_{1}$. 

We will use a recursion of depth $r$,
  a constant to be determined later.
We let $A_{0} = A$ denote the initial matrix.
We let $B_{i+1}$ denote the preconditioner
  for $A_{i}$,
 $L_{i} C_{i} L_{i}^{T}$ be the partial $LDL^{T}$
  factorization of $B_{i}$ in trim order,
  and $C_{i} = [D_{i}, 0; 0, A_{i}]$.
To analyze the algorithm, we must determine
  the relative error $\epsilon_{i}$ to which
  we will solve the systems in $A_{i}$.
The bound we apply is derived from the following lemma,
  which we derive from a result of
  Golub and Overton~\cite{GolubOverton}.

\begin{lemma}[Preconditioned Inexact Chebyshev Method]
\label{lem:GolubOverton}
Let $A$ and $B$ be Laplacian matrices satisfying
  $\sigma (B, A) \geq 1$.
Let $\xx$ be the solution to $A\xx = \bb$.
If, in each iteration of the preconditioned  Chebyshev Method,
  a vector $\zz_{k}$ is returned satisfying
\[
  B \zz_{k} = \rr_{k} + \qq_{k}, \mbox{ where $\norm{\qq_{k}} 
  \leq \delta \norm{\rr_{k}}$},
\]
where 
$\delta 
\leq 
\left(128 \sqrt{\kappa_{f} (B)} \sigma (A,B) \right)^{-1}$,
then the $k$-th iterate, $\xx_{k}$, output by the algorithm will satisfy
\[
  \norm{\xx-\xx_{k}}
\leq 
6 \cdot 2^{- k / \sqrt{\kappa_{f} (A,B)}}
  \kappa_{f} (A) \sqrt{\kappa_{f} (B)} \norm{\xx}
.
\]
\end{lemma}


Our main theorem is:

\begin{theorem}[Recursive]\label{thm:recursive}
Let $A$ be an $n$-by-$n$ PSDDD matrix with $m$ 
  non-zero entries.
Using the recursive algorithm, one can solve the system
  $A\xx = \bb $ to relative accuracy $\epsilon $ in
  time 
\[
\bigO{m^{1.31+o (1)} (\log (\epsilon^{-1} ) 
   \log (n \kappa (A)))^{O (1)}}.
\]
Moreover, 
  if the graph of $A$ does not contain a
  minor isomorphic to the complete graph on $m^{\theta }$
           vertices, or has genus at most $m^{2\theta }$, 
  or is the Gremban cover of such a graph,
  then the exponent of $m$ can be reduced to $1 + 5 \theta + o (1)$.
\end{theorem}

We note that  if $G (A)$ is planar, then the algorithm
  take time nearly linear in $m$.

The following two lemmas allow us to bound the accuracy
  of the solutions to systems in $B_{i}$ in terms  
  of the accuracy of the solutions to the corresponding systems
  in $A_{i}$.

\begin{lemma}\label{lem:conditionL}
Let $L C L^{T}$ be a partial $LDL^{T}$-decomposition
  of a symmetric diagonally dominant matrix.
Then,
\[
 \kappa (L) \leq 2 n^{3/2}.
\]
\end{lemma}

\begin{proof}
As $L$ is column diagonally-dominant
  and has 1s on its diagonal,
  $\norm{L}_{1} \leq 2$;
  so, $\norm{L} \leq 2 \sqrt{n}$.
By a result of Malyshev~\cite[Lemma~1]{Malyshev},
  $\norm{L^{-1}} \leq n$
  (also see Pe{\~n}a~\cite{Pena}).
\end{proof}

\begin{lemma}\label{lem:partialLDL}
Let $B$ be a Laplacian matrix, let
  $L C L^{T}$ be the partial $LDL^{T}$-factorization obtained
  by eliminating vertices of $B$ in the trim order.
Then, $\kappa (C) \leq \kappa (B)$.
\end{lemma}
\begin{proof}
We recall that $C$ has form
\[
  \left(\begin{array}{ll} 
     D & 0 \\
     0 & A_{1}
   \end{array}\right).
\]
The factor $A_{1}$ is identical to that obtained from partial
  Cholesky factorization, so $\kappa (A_{1}) \leq \kappa (B)$
  follows from Lemma~\ref{lem:partialCholesky}.
To now bound $\kappa (C)$, we need merely show that each entry
  of $D$ lies between the smallest and largest non-zero eigenvalues of
  $B$.
This follows from the facts that the $i$th diagonal of $D$ equals
  the value of the diagonal of the corresponding vertex
  in the lower factor right before it is eliminated,
  this value lies between the smallest and largest non-zero elements
  of the corresponding factor, and by Lemma~\ref{lem:partialCholesky},
  these lie between the largest and smallest non-zero eigenvalues of $B$.
\end{proof}

\begin{lemma}\label{lem:errors}
Let $B$ be a Laplacian matrix and let
  $L C L^{T}$ be the partial $LDL^{T}$-factorization obtained
  by eliminating vertices of $B$ in the trim order.
For any $\cc \in \Span{B}$,
  let $\sss$ be the solution to $C \sss  = L^{-1}\cc $
  and let $\sst$ satisfy
  $\norm{\sss -\sst} \leq \epsilon \norm{\sss}$.
Let $\approximation{\yy}$
  be the solution to $L^{T}\approximation{\yy} 
 = \sst$.
Then
\[
\norm{\cc - B \yyt } \leq \epsilon\kappa (L)\kappa (C)  \norm{\cc }.
\qedhere
\]
\end{lemma}

\begin{proof}[Proof of Lemma~\ref{lem:errors}]
First, note that $\cc - B \yyt = L C (\sss- \sst)$
  and $\cc = L C \sss $.
Moreover, $L^{-1}\cc $ must lie in $\Span{C}$.
Thus, $\norm{C \sss} \geq \lambda_{2} (C) \norm{\sss }$, and
  so 
\[
  \norm{C (\sss - \sst)} \leq \epsilon \kappa_{f} (C) \norm{C \sss }.
\]
As $L$ is non-degenerate, we may conclude
\[
  \norm{L C (\sss - \sst)} 
 \leq \epsilon \kappa (L)  \kappa_{f} (C) \norm{L C \sss }.
\]
\end{proof}

\begin{proof}[Proof of Theorem~\ref{thm:recursive}]
For $A_{0}, \ldots , A_{r}$,
  $B_{1}, \ldots , B_{r}$,
  $C_{1}, \ldots , C_{r}$,
  and $L_{1}, \ldots , L_{r}$ as defined above,
  we can apply 
  Lemma~\ref{lem:partialLDL} and Theorem~\ref{thm:main}
  to show:
\begin{itemize}
\item $\kappa _{f} (A_{i}) \leq \kappa _{f} (B_{i})
  \leq m^{i (1+o (1))} \kappa _{f} (A)$,
\item $\kappa _{f} (B_{i}) \leq m^{1+o (1)} \kappa _{f} (A_{i-1})
   \leq m^{i (1+o (1))} \kappa _{f} (A)$
\end{itemize}

In the recursive algorithm we will solve systems
  in $A_{i}$, for $i \geq 1$, to accuracy
\[
  \epsilon_{i} = 
  \left(128 m^{i (1+o (1))} (2 n^{3/2} \kappa (A)) \right)^{-1}.
\]
By Lemma~\ref{lem:errors} and the above bounds, 
  we then obtain solutions to
  the systems in $B_{i}$ to sufficient accuracy
  to apply Lemma~\ref{lem:GolubOverton}.

Let $m_{i}$ be the number of edges of $A_{i}$.
When constructing the preconditioner, we  
  set $t_{i} = (m_{i})^{\gamma }$, for a
  $\gamma$ to be chosen later.
Thus, by Theorem~\ref{thm:main}
  and Proposition~\ref{pro:trim},
  $m_{i} \leq m^{(2\gamma)^{i}}$,
  and
  $\kappa_{f} (A_{i}, B_{i+1}) = m^{(2\gamma)^{i} (1-\gamma) + o (1)}$.

We now prove by induction that the running time
  of the algorithm obtained from a depth $r$
  recursion is
\[
  \bigO{m^{\beta_{r} + o (1)} \left(r \log (n \kappa (A) \right)^{r}},
\mbox{ where }
\]
\[
 \beta_{r} \defeq 
  (\frac{1-\gamma}{2})
  \sum_{i=1}^{r} (2 \gamma)^{i-1} 
 +  2 (2\gamma)^{r},
\]
and $\gamma \defeq (3 - \sqrt{5})/2$.
In the limit,
  $\beta_{r}$ approaches
  $\beta_{\infty} \defeq (3 + \sqrt {5})/4$ from above.
The base case, $r = 1$,
  follows from Theorem~\ref{thm:oneShot}.

The preprocessing time is negligible as
  the partial Cholesky factorizations used to produce
  the $C_{i}$ take linear time,
  and the full Cholesky factorization is only
  performed on $A^{r}$.

Thus, the running time is bounded by the 
  iterations.
The induction follows by 
  observing that the iteration time
  is
$m^{\frac{1-\gamma}{2} + o (1)} 
  \left(m + m_{1}^{\beta_{r-1}} \right)
  \log \left(\kappa (A_{r}) \kappa (B_{r}) / \epsilon_{r} \right)
$, which proves the inductive hypothesis because
 $ m_{1}^{\beta_{r-1}} > m$.
As $1.31 > \beta_{\infty}$, there exists an
  $r$ for which $\beta_{r} < 1.31$.

When the graph of $A$ does not contain a
  $K_{m^{\theta}}$ minor or has genus at
  most $m^{2\theta }$,
  we apply a similar analysis.
In this case, we have
  $m_{i} \leq m_{i-1} m^{\theta + o (1)}$.
Otherwise, our proof is similar, except that
  we set
  $\gamma = (3 - \theta - \sqrt{1 + 6 \theta + \theta^{2}})/2$,
  and obtain
  $\beta_{\infty} = (3 + \theta + \sqrt{1 + 6 \theta + \theta^{2}})/4$,
  and note that
  $\beta_{\infty} \leq 1 + 5 \theta$.
\end{proof}
